\documentclass{ws-procs9x6-cpt19}
\begin{document}

\newcommand{\refeq}[1]{(\ref{#1})}
\def\etal {{\it et al.}}

\title{The MICROSCOPE Space Mission and Lorentz Violation
}

\author{Geoffrey Mo,$^1$
  H\'el\`ene Pihan-Le Bars,$^2$
  Quentin G.\ Bailey,$^3$
  Christine Guerlin,$^{2,4}$  \\
  Jay D.\ Tasson,$^1$ and 
  Peter Wolf$^2$}

\address{$^1$Department of Physics and Astronomy, Carleton College,\\
  Northfield, MN 55057, USA}

\address{$^2$SYRTE, Observatoire de Paris, Universit\'e PSL, CNRS,\\
  Sorbonne Universit\'e, LNE, 75014 Paris, France}

\address{$^3$Department of Physics and Astronomy, 
  Embry-Riddle Aeronautical University,\\
Prescott, AZ 86301, USA}

\address{$^4$Laboratoire Kastler Brossel, ENS-Université PSL, CNRS, Sorbonne Universit\'e, Coll\`ege de France, 24 rue Lhomond, 75005 Paris, France}

\begin{abstract}
  In this contribution to the CPT'19 proceedings, 
  we summarize efforts that use data from the MICROSCOPE space mission
  to search for Lorentz violation in the Standard-Model Extension.
\end{abstract}

\bodymatter

\section{The SME and the Weak Equivalence Principle}

Data from the MICROSCOPE space mission 
can be used to search for Lorentz violation
within the field-theoretic framework of the Standard-Model Extension (SME).\cite{us}
The SME can roughly be thought of as a series expansion about known physics
at the level of the action
that forms a broad and general framework for tests of Lorentz symmetry.\cite{sme,rev}
Terms in the SME action
are constructed from Lorentz-violating operators
along with coefficients for Lorentz violation
that characterize the amount of Lorentz violation in the theory.
In general, 
the coefficients for Lorentz violation
are particle-species dependent
such that in the fermion sector couplings to Lorentz violation
may differ for protons, neutrons, and electrons.
When couplings to gravity are considered in the fermion sector,
this species dependence 
leads to effective Weak Equivalence Principle (WEP) violations.\cite{lvgap}

In the present context,
we consider a single coefficient field $(a_{\rm eff})_\mu$
in the classical point-particle limit,
where the matter-sector action is\cite{lvgap}
\begin{equation}
  S^{\rm B}_U \approx \int{}d\lambda{}\big(-m^{\rm B}\sqrt{-g_{\mu\nu}u^\mu u^\nu} - (a^{\rm B}_{\rm eff})_\mu u^\mu\big),
  \label{eq:action}
\end{equation}
for a body B of mass $m^{\rm B}$
with the post-Newtonian metric $g_{\mu\nu}$ and four-velocity $u^\mu$.
In spontaneous Lorentz-violation models,
$(a_{\rm eff})_\mu$ develops a vacuum expectation value $(\bar a_{\rm eff})_\mu$.
An additional freedom associated with a coupling constant 
in models of spontaneous breaking is characterized by~$\alpha$.\cite{lvgap}

When these coefficients are taken into account in a Newtonian example 
via the action (\ref{eq:action}),
the equations of motion for bodies become
\begin{equation}
m^{\rm B} \vec {\rm a}   = ( m^{\rm B} + 2\alpha(\bar{a}_{\rm eff}^{\rm B})_t) \vec g,
  \label{eq:onBody}
\end{equation}
where $\vec {\rm a}$ is the acceleration,
$\vec g$ is the Newtonian gravitational field,
and $(\bar{a}_{\rm eff}^{\rm B})_\mu{} = \sum_w N_{\rm B}^w(\bar{a}^w_{\rm eff})_\mu{}$.\cite{lvgap}
Here, 
the sum is over particle species $w=$ proton, neutron, electron,
and $N_{\rm B}^w$ is the number of particles of type $w$ in the body.

The WEP states 
that the gravitational mass $m_{\rm grav}$
is equal to the inertial mass $m_{\rm inert}$.\cite{will}
In other words,
\begin{equation}
  m_{\rm inert} \vec {\rm a}= m_{\rm grav} \vec g
  \label{eq:inertial}
\end{equation}
can be rewritten as
$ \vec {\rm a}= \vec g$.
Hence,
the signal in a WEP experiment
is a relative acceleration
of a pair of co-located bodies 1 and 2 
in free fall in a gravitational field.
Traditional constraints on Lorentz-invariant WEP violations
have been quantified by the E\"otv\"os parameter
$  \delta{}(A,B) = 2({\rm a}_1 - {\rm a}_2)/({\rm a}_1+{\rm a}_2)$,
where ${\rm a}_B$ is the free-fall acceleration of the body.

Comparison of Eqs.\ (\ref{eq:onBody}) and (\ref{eq:inertial})
reveals the effective WEP violation induced by $(\bar a_{\rm eff})_\mu$.
However,
the Lorentz-violation signal is more complicated 
and cannot be characterized by a single parameter
such as $\delta$.
This can be seen already from the Newtonian result,
where the time component of the coefficient for Lorentz violation
will mix with the spatial components under a time-dependent boost,
such as the boost of the Earth around the Sun,
yielding a time-dependent WEP signal 
involving multiple components of $(\bar a_{\rm eff})_\mu$.
When the relative free-fall rate for a pair of electrically neutral bodies 1 and 2 
is considered,
the signals are proportional to\cite{lvgap}
\begin{equation}
  \sum_w\left(\frac{N_1^w}{m^1} - \frac{N_2^w}{m^2}\right)(\bar{a}_{\rm eff}^w)_\mu
  \approx   \frac{N_1^n N_2^p-N_1^p N_2^n}{m^1 m^2} m^n  (\bar{a}_{\rm eff}^{n-e-p})_\mu \equiv A (\bar{a}_{\rm eff}^{n-e-p})_\mu,
  \label{species}
\end{equation}
where $(\bar{a}_{\rm eff}^{n-e-p})_\mu \equiv (\bar{a}_{\rm eff}^n)_\mu - (\bar{a}_{\rm eff}^e)_\mu - (\bar{a}_{\rm eff}^p)_\mu$.

\section{MICROSCOPE space mission}
The MICROSCOPE space mission is a French CNES microsatellite
designed to test the WEP 
with the best-ever precision of one part in $10^{15}$.\cite{microscopePlan}
It was launched in 2016 
and completed its data-taking in 2018. 
It contains an instrument 
composed of two concentric cylindrical test masses
made of the platinum alloy Pt:Rh (90:10) 
and the titanium alloy Ti:Al:V (90:6:4) 
for the inner and outer mass, 
respectively.
With these measurements, 
the MICROSCOPE team found a constraint 
on the E\"otv\"os parameter of \mbox{$\delta$(Ti, Pt) = 
[1 $\pm$ 9 (stat) $\pm$ 9 (syst)] $\times$ $10^{-15}$},
representing an improvement of over an order of magnitude 
over the previous best limits.\cite{microscopeResult}

A subset of the Lorentz-violation signals 
arise at  frequencies different from the WEP signal 
and may remain hidden in the analysis aimed at the traditional WEP.
Hence, 
a separate analysis has been performed
to search for signals at the additional frequencies associated with 
$(\bar{a}_{\rm eff}^{n-e-p})_\mu$.\cite{us}
Constraints on coefficients were analyzed globally
treating all four components together 
as well as for ``maximal reach,"\cite{gravimeter} 
where only one coefficient was assumed to be nonzero at a time.
With the alloys used for the test bodies,
the parameter $A$ in Eq.\ (\ref{species}) is about $0.06\,$GeV$^{-1}$.
With this, 
improvements on prior sensitivities\cite{datatables} 
of about one to two orders of magnitude were achieved.\cite{us}

\section*{Acknowledgments}
G.M.\ is grateful for support from the Carleton College Towsley fund,
and Q.G.B.\ was supported by NSF grant 1806871.

\end{document}